\documentclass[a4paper,12pt]{article}
\usepackage{float}
\usepackage{tabularx}
\usepackage{amsmath}
\usepackage{amsfonts}
\usepackage{adjustbox}
\usepackage{multirow}
\usepackage{color}
\usepackage{longtable}
\usepackage{pdflscape}
\usepackage{rotating}
\usepackage{xcolor}

\usepackage{xcolor}
\usepackage[colorlinks = true,
            linkcolor = blue,
            urlcolor  = blue,
            citecolor = blue,
            anchorcolor = blue]{hyperref}
\usepackage[margin=1in,footskip=0.25in]{geometry}
\title{Isospectral Potentials and Quantum Mechanical Functions for Neutron-Neutron Scattering }
\author{Anil Khachi$^{1}$\\\\
$^{1}$ Department of Physics\\ St. Bedes College, 171002, \\Himachal Pradesh, Bharat (India)}
\begin{document}
\maketitle
\abstract{\noindent In this paper we have constructed inverse isospectral potentials for $^1S_0-nn$ state by fitting the experimental SPS using Variational Monte-Carlo technique in tandem with PFM technique. The isospectral potentials are obtained such that the cost measure i.e., mean absolute error (MAE) between the obtained and experimental SPS are less than 1 and the parameters give the low energy scattering parameters (`$a$' and `$r_e$') very close to the experimental values. The S-channel \textit{SPS} for $^1S_0-nn$ have been obtained, with a MAE with respect to experimental data for lab energies up to 350 MeV, to less than 1.}\\
{\textbf{keywords:}  nn-scattering, Inverse potential, Phase function method, Variational Monte-Carlo, Morse potential, Amplitude, wavefunction}

\section{Introduction}
The experimental setup in neutron-neutron scattering consists of a neutron beam generated by $^3H(d,n)^4He$ reaction. The target consists of heavy water enriched in $D_2O$. In total the nd-breakup reaction i.e., $n+d(deuterium) \rightarrow p+n+n$ is the reaction leading to the appearance of interacting two neutrons in the final state. In addition to n-d, another reaction i.e., $\pi^+$ $\rightarrow$ $\gamma+n+n$ helps in determinining the scattering length for \textit{nn} reaction. \\

Neutron-neutron interaction is important because the reaction helps in the investigation of charge idependence which has been fundamentally interesting to the nuclear community. It is well known that to investigate the charge independence in nuclear forces it is quite sufficient to compare the \textit{np} and \textit{pp} interactions. It is also known that to get more accurate information about these forces one needs to get precise low energy scattering data. Under such circumstances only S- waves are important. The investigation of low energy scattering parameters is directly related to test the hypothesis of charge independence and charge symmetry. Both theoreticians and experimentalists are keen to precisly determine the low energy parameters for \textit{np, pp} and \textit{nn} interactions.

The nucleon-nucleon interactions at low energies are usually expressend in terms of effective range ($r_e$) and scattering length (\textit{a}). These parameters provides information on the charge idependence of the nuclear forces. There have been various arguments regarding the possibility of violation of charge independence and charge symmentry some of them are discussed here:

\begin{itemize}
\item Darewych \textit{et al.} \cite{1} found that there was substantial difference between the \textit{np} and \textit{nn} potential curves with $^1S_0-nn$ ($V_0$=40.38 MeV) and for $^1S_0-np$ ($V_0$=61.99 MeV) hence observed ``violation of charge independence". The violation was credited to the unavailability of high energy \textit{nn} data. They used high energy \textit{np} and \textit{pp} data for fitting the \textit{nn} SPS curve.

\item Henley and others \cite{2} have discussed about the small departure from charge symmetry and charge indpendence. The departure can be originated by \textit{em} forces and is related to the basic understanding nuclear interaction and is ultimately connected with the fundamental particle principles of elementary physics. Since departures have been observed to be very small it is imperative to acquire precise data on the nucleon nucleon interaction.

\item Babenko and Petrov \cite{3} observed from a comparison of the low-energy parameters for the \textit{np} system and their counterparts \textit{pp} and \textit{nn} systems that the charge dependence of nuclear forces is violated, which is associated with the mass difference between the charged and neutral pions.

\item The values of scattering length and effective range are sensitive to any small change in \textit{nn} potentials. As stated by E.S. Konobeevski \textit{et al.} \cite{4}. A change of 7\% in V(r) may lead to change of 20-30\% in the scattering length calculation.
\end{itemize} 

The above mentioned points are either related to the precise knowledge of data or related to the fundamental knowledge of elementary physics. Our main aim in this paper would be to use the adapted nn data by Wiringa \textit{et al.} \cite{5} and obtain the interaction potentials by GOA and CDA techniques developed by our group \cite{6}\cite{7}\cite{8}\cite{9}. 
\begin{itemize}
\item By GOA we will be running 3 parameters ($V_0, r_m \& a_m$) for all 11 SPS data points at once and obtain one set of globally optimized model parameters. This is shown by solid blue line in Figure \ref{SPS_POT}.
\item While by CDA we will be running 3 parameters for only 3 SPS data points. The CDA results creates a set of isospectral potential shown in Figure \ref{SPS_POT} as yellow bar.The details of the CDA procedure have been provided in the Appendix section \ref{Appendix}.
\end{itemize}

\section{Methodology:}
The Morse function \cite{10} is given by:
\begin{equation}
V_\text{Morse}(r) = V_0\left(e^{-2(r-r_m)/a_m}-2e^{-(r-r_m)/a_m}\right)
\label{eq1}
\end{equation} 
where the model parameters $V_0$, $r_m$ and $a_m$ reflect strength of interaction, equilibrium distance at which maximum attraction is felt (equilibrium separation) and range parameter respectively. It has all the interesting features that are observed in any typical scattering experiment, such as, strong repulsion at short distances, maximum attraction at an equilibrium distance $r_m$, followed by a quickly decaying tail at large distances. It can also be observed that realistic N-N interaction potentials, of Argonne v18 and Reid93 \cite{8}, for S-sates resemble that of a Morse function. Also, the phenomenological Malfeit-Tjohn potential \cite{11} has a similar shape. Further, the analytical solution for the \textit{TISE} with Morse potential interaction has already been solved for bound state energies ($E < 0$) \cite{12}. 
Hence, it can be considered as a good choice for modeling the interaction between any two scattering particles.\\

\subsection{Phase Function Method:}  
The Schr$\ddot{o}$dinger wave equation for a spinless particle with energy E and orbital angular momentum $\ell$ undergoing scattering is given by
\begin{equation}
\frac{\hbar^2}{2\mu} \bigg[\frac{d^2}{dr^2}+\big(k^2-\ell(\ell+1)/r^2\big)\bigg]u_{\ell}(k,r)=V(r)u_{\ell}(k,r)
\label{Scheq}
\end{equation}
Where $k=\sqrt{E/(\hbar^2/2\mu)}$.

Second order differential equation  Eq.\ref{Scheq} has been transformed to the first order non-homogeneous differential equation of Riccati type \cite{13,14} given by following equation:  
\begin{equation}
\delta_{\ell}'(k,r)=-\frac{V(r)}{k}\bigg[\cos(\delta_\ell(k,r))\hat{j}_{\ell}(kr)-\sin(\delta_\ell(k,r))\hat{\eta}_{\ell}(kr)\bigg]^2
\label{PFMeqn}
\end{equation}
 
Here in Eq. \ref{PFMeqn} prime denotes differentiation of phase shift with respect to distance and the Riccati Hankel function of first kind is related to $\hat{j_{\ell}}(kr)$ and $\hat{\eta_{\ell}}(kr)$ by $\hat{h}_{\ell}(r)=-\hat{\eta}_{\ell}(r)+\textit{i}~ \hat{j}_{\ell}(r)$ . In integral form the above equation can be simply written as:
\begin{equation}
\delta(k,r)=\frac{-1}{k}\int_{0}^{r}{V(r)}\bigg[\cos(\delta_{\ell}(k,r))\hat{j_{\ell}}(kr)-\sin(\delta_{\ell}(k,r))\hat{\eta_{\ell}}(kr)\bigg]^2 dr
\end{equation}
 
Eq.\ref{PFMeqn} is numerically solved using Runge-Kutta 5$^{th}$ order (RK-5) method with initial condition $\delta_{\ell}(0) = 0$. For $\ell = 0$, the Riccati-Bessel and Riccati-Neumann functions $\hat{j}_0$ and $\hat{\eta}_0$ get simplified as $\sin(kr)$ and $-\cos(kr)$, so Eq.\ref{PFMeqn}, for $\ell = 0$ becomes 
\begin{equation}
\delta'_0(k,r)=-\frac{V(r)}{k}\sin^2[kr+\delta_0(k,r)]
\end{equation}
The equation for amplitude function \cite{15} with initial condition is obtained in the form

\begin{equation}
\begin{aligned}
    A_{\ell}^{\prime}(r) = &-\frac{A_{\ell} V(r)}{k} \left[\cos (\delta_\ell(k,r)) \hat{j}_{\ell}(kr)-\sin (\delta_\ell(k,r)) \hat{\eta}_{\ell}(kr)\right] \\
    &\times\left[\sin (\delta_\ell(k,r))( \hat{j}_{\ell}(kr)+\cos (\delta_\ell(k,r)) \hat{\eta}_{\ell}(kr)\right]
\end{aligned}
\end{equation}
\newline
also the equation to obtained wavefunction \cite{15} is 
\begin{equation}
    u_{\ell}(r)=A_{\ell}(r)\left[\cos (\delta_\ell(k,r)) \hat{j}_{\ell}(k r)-\sin (\delta_\ell(k,r)) \hat{\eta}_{\ell}(k r)\right]
\end{equation}
In above equation the function $\delta_0(k,r)$ was termed ``Phase function" by Morse and Allis \cite{16}. The significant advantage of \textit{PFM} method is that, the phase-shifts are directly expressed in terms of the potential and have no relation to the wavefunction. This has been utilised in this paper to obtain inverse potentials in an innovative way by implementing a modified \textit{VMC} in tandem with \textit{PFM}. The technique optimizes the model parameters of the potential to obtain the best match with respect to the experimental SPS values. Also, rather than solving the second order Schr$\ddot{o}$dinger equation, we only need to solve the first order non-homogeneous differential equation whose asymptotic value gives directly the value of \textit{SPS}.
\subsection{Optimization of Morse function model parameters using VMC:}
Typically, \textit{VMC} is utilized for obtaining the ground state energy for a given potential. The method initiates with a trial wavefunction, which is varied at a random location by a random amount in the Monte-Carlo sense. Then, the energy is determined using the newly obtained wavefunction and variational principle is applied. This process is done iteratively till one converges to the ground state. Here, we consider to vary the potential instead of wavefunction and minimise variance with respect to experimental data, as follows:\\
\textbf{Initialisation step:} To begin the optimisation procedure, Morse parameters $V_0$, $r_m$ and $a_m$ are given some initial values. The phase equation is integrated using RK-5 method for different values of $k$, a function of lab energies E, to obtain the simulated \textit{SPS}, say \textit{$\delta^{sim}_k$}. The mean percentage error (\textit{MPE}) has been determined with respect to the SPS analysis data of Wiringa \textit{et al.}, \cite{5}, say \textit{$\delta^{exp}_k$}, as\\

\begin{equation}
MPE = \frac{1}{N}\sum_{i=1}^N \frac{\lvert\delta^{exp}_k-\delta^{sim}_k\rvert}{\lvert\delta^{exp}_k\rvert} \times 100
\end{equation}
This is named as $MPE_{old}$ and is also assigned to $MPE_{min}$.\\
\textbf{Monte-Carlo step:} A random number $r$, generated in an interval [-I, I], is added to one of the parameters, say $V_{0new} = V_0 + r$.\\
\textbf{PFM step:} Again, the phase equation is integrated with new set of parameters $V_{0new}$, $r_m$ and $a_m$ to obtain new set of simulated scattering phase shifts (\textit{SPS}), say \textit{$\delta^{sim-new}_k$}, using which $MPE_{new}$ is determined.\\
\textbf{Variational step:} If $MPE_{new}<MPE_{old}$, then $V_0=V_{0new}$, $MPE_{min}=MPE_{new}$, else old values are retained. \\

The final three steps are repeated for each of the parameters to complete one iteration. The size of interval is reduced after a certain number of iterations, if there is no significant reduction in $MPE_{min}$. The process is completed when $MPE_{min}$ does not change any further, that is, convergence is reached.
\section{Singlet Pseudo Bound S-wave Energy for \textit{nn}}
We know:
\begin{equation}
    k\cot(\delta)=-\frac{1}{a}+\frac{1}{2}kr_e^2
    \label{kcot}
\end{equation}
where \textit{a} and $r_e$ are scattering length  and effective range. If use is made of the approximation specified by Eq. \ref{kcot}, the S matrix can be written in the form \cite{3}:
\begin{equation}
    S(k)=\left(\frac{k+i\alpha}{k-i\alpha}\right) \left(\frac{k+i\beta}{k-i\beta}\right)
    \label{Smatrix}
\end{equation}
Where S(k) is related to scattering length \textit{a} and effective range $r_e$ by following relations
\begin{equation}
    \alpha=\frac{1}{r_e}\bigg[1-\big(1-\frac{2r_e}{a}\big)^{1/2}\bigg]
    \label{Alpha}
\end{equation}
\begin{equation}
    \beta=\frac{1}{r_e}\bigg[1+\big(1-\frac{2r_e}{a}\big)^{1/2}\bigg]
\end{equation}

The S matrix in Eq. \ref{Smatrix} has two poles in the complex plane of the wave number k.

The first pole $i\beta$ ($\beta > 0$),
situated in the upper half-plane of k, is the well-known redundant pole of the S matrix. The second pole $i\alpha$ , situated in the lower half-plane of k, corresponds to a virtual ($i\alpha < 0$) state of the two-nucleon system at the energy.
\begin{equation}
    \epsilon=\frac{\hbar^2\alpha^2}{2m}
    \label{Senergy}
\end{equation}
Using the obtained values for $a=-16.49(1.56)$ and $r_e=2.56(0.02)$ in Eq. \ref{Alpha} and then substituting the value in Eq. \ref{Senergy} the energy of the virtual $^1S_0-nn$ state comes out to be 0.135(22) MeV in comparison with $0.1293(158)$ given in \cite{3}.

\section{Results and Discussion}
\begin{figure*}
\centering
{\includegraphics[scale=0.54,angle=0]{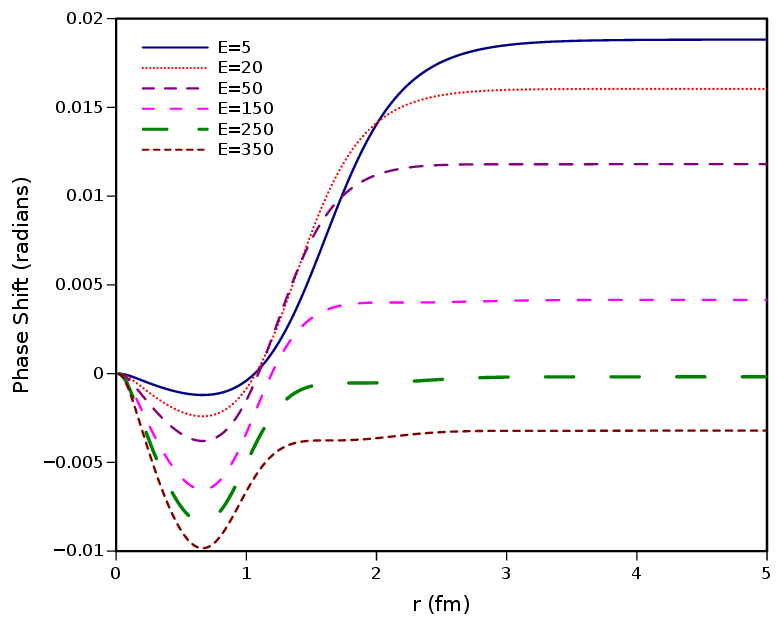}}
{\includegraphics[scale=0.5,angle=0]{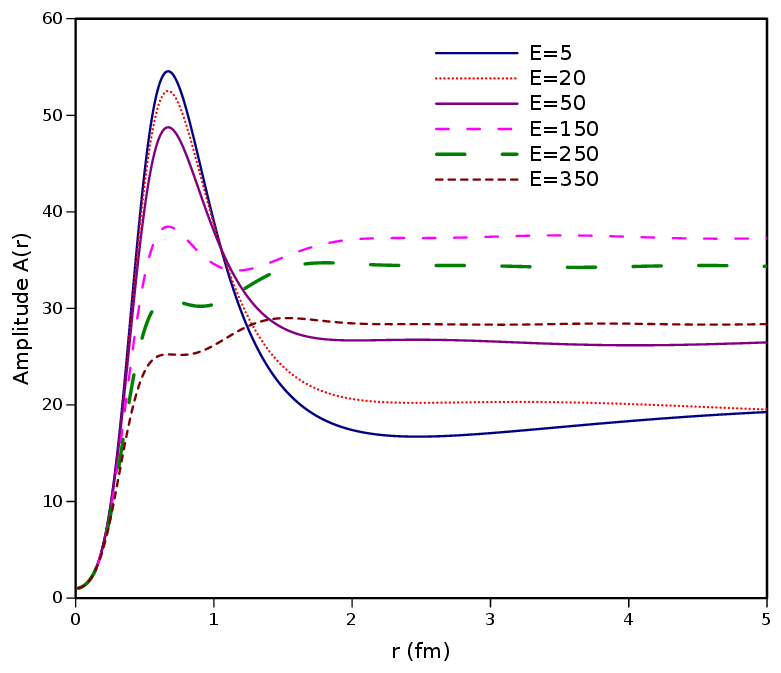}}
{\includegraphics[scale=0.8,angle=0]{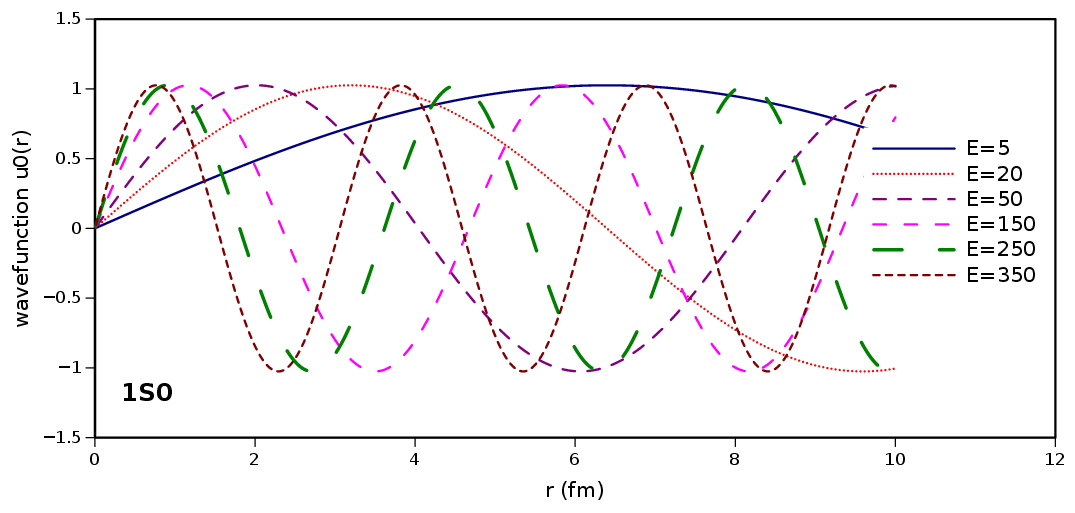}}
\caption{$^1S_0-nn$ SPS, amplitude and wavefunction function vs. distance r(fm) for $^1S_0-nn$ state.}
\label{S_r}
\end{figure*}

At laboratory energies of nn scattering larger that $E_{\ell ab} = 250 MeV$ the measured scattering phase-shift in the $^1S_0$-wave interaction channel becomes negative, i.e., the interaction becomes repulsive. The fits to SPS are improved to reproduce the experimentally observed scattering length and effective range. To reproduce the observed negative S-wave phase shifts at higher energies with a static potential it is necessary to incorporate a repulsion into the potential. Here the Morse potential serves as a far more realistic potential in comparison to square well, exponential, Gaussian, or Hulthen potential used in the earlier studies. In addition to SPS and pseudo bound state energy for the \textit{nn} system we have also obtained important quantum mechanical functions like (i) SPS vs. r(fm) (ii) Amplitude A(r) vs r(fm) and Wavefunction $u_0(r)$ for energy E=[5, 20, 50, 150, 250, 350]. These functions are shown in figure \ref{S_r} respectively.
\begin{table}[]
\centering
\caption{Optimized parameters for $^1S_0-nn$ states using GOA and CDA. In later case, parameter values consisting of extreme depths are shown. Scattering length ($a$ in fm) and effective range ($r_e$ in fm) obtained, using SPS determined from these optimized parameters, are shown with experimental values (bold) \cite{5} in curly  brackets. Virtual $^1S_0$ state energy is taken from Petrov \textit{et al.}} 
\scalebox{0.8}{
\setlength{\tabcolsep}{8pt} 
\renewcommand{\arraystretch}{1.4} 
\begin{tabular}{cccccc} 
\hline\hline
Analysis                                 & $[V_0$, $r_m$, $a_m]$     & MAE & $a (fm)$ & $r_e (fm)$  & $E_{nn}$ \\ 
\hline
\multirow{1}{*}{GOA}                     & {[}67.523, 0.925, 0.376]  & 0.5 & --17.64$\{$\textbf{-18.5(0.4)}$\}$ & 2.58$\{$\textbf{2.8(0.11)}$\}$ & 0.117\{\textbf{0.1293(158)}\}\\

\hline

\multirow{2}{*}{ CDA  }  & {[}74.427,0.915,0.359]  & 0.7 & \multirow{2}{*}{-16.49(1.56)}&\multirow{2}{*}{2.56(0.02)} &  \multirow{2}{*}{0.135(22)\{\textbf{0.1293(158)}\}}\\

                        \vspace{0.05cm}

                        &                         {[}65.191,0.915,0.381]  & 0.9   \\
\hline
\end{tabular}
}
\label{scatparam}
\end{table}

\begin{figure*}
\centering
{\includegraphics[scale=0.42,angle=270]{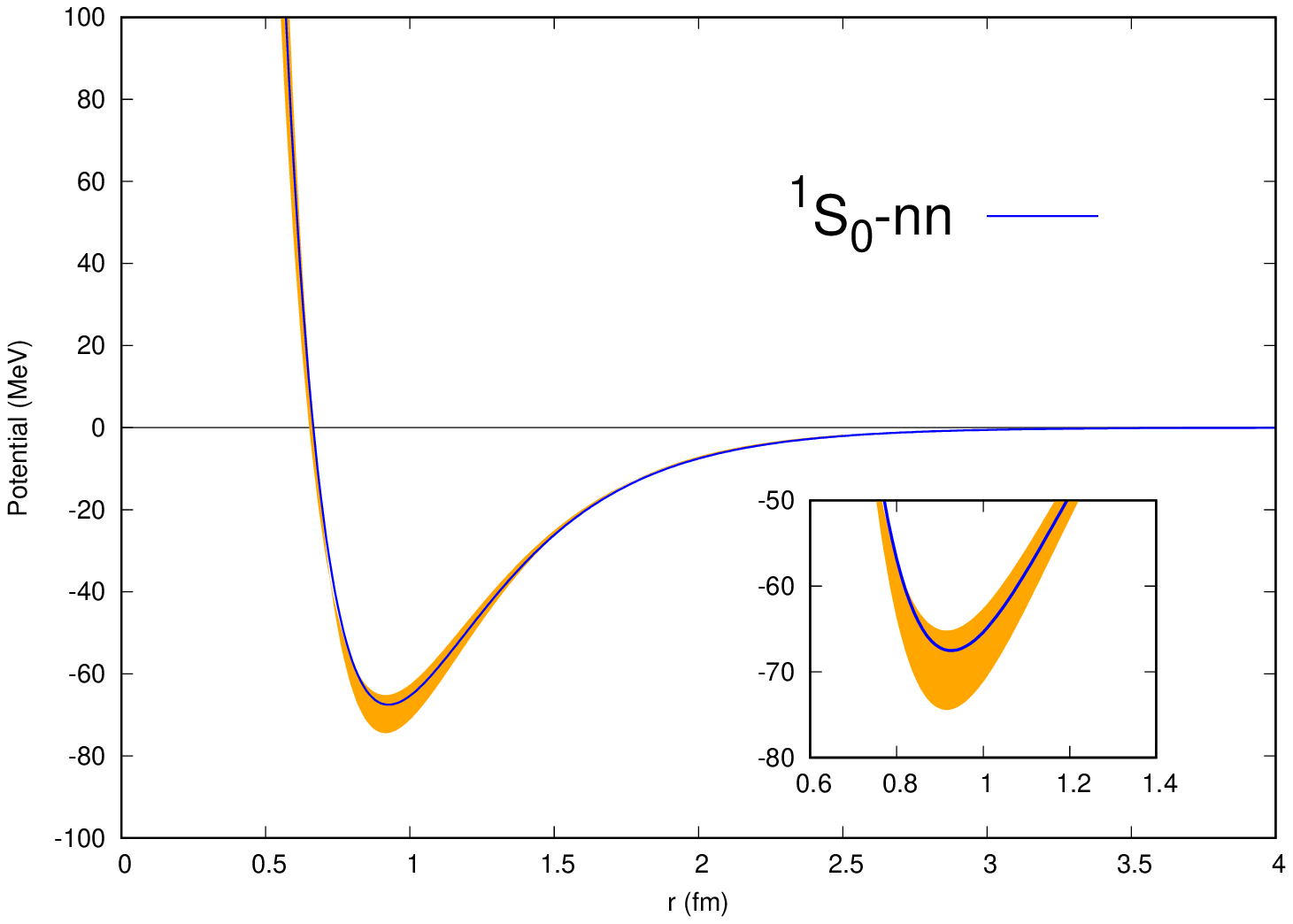}}
\caption{$^1S_0-nn$ potentilas obtained through GOA (solid blue line) and CDA (yellow ribbon) analysis.}
\label{SPS_POT}
\end{figure*}

\begin{figure*}
\centering
{\includegraphics[scale=0.42,angle=270]{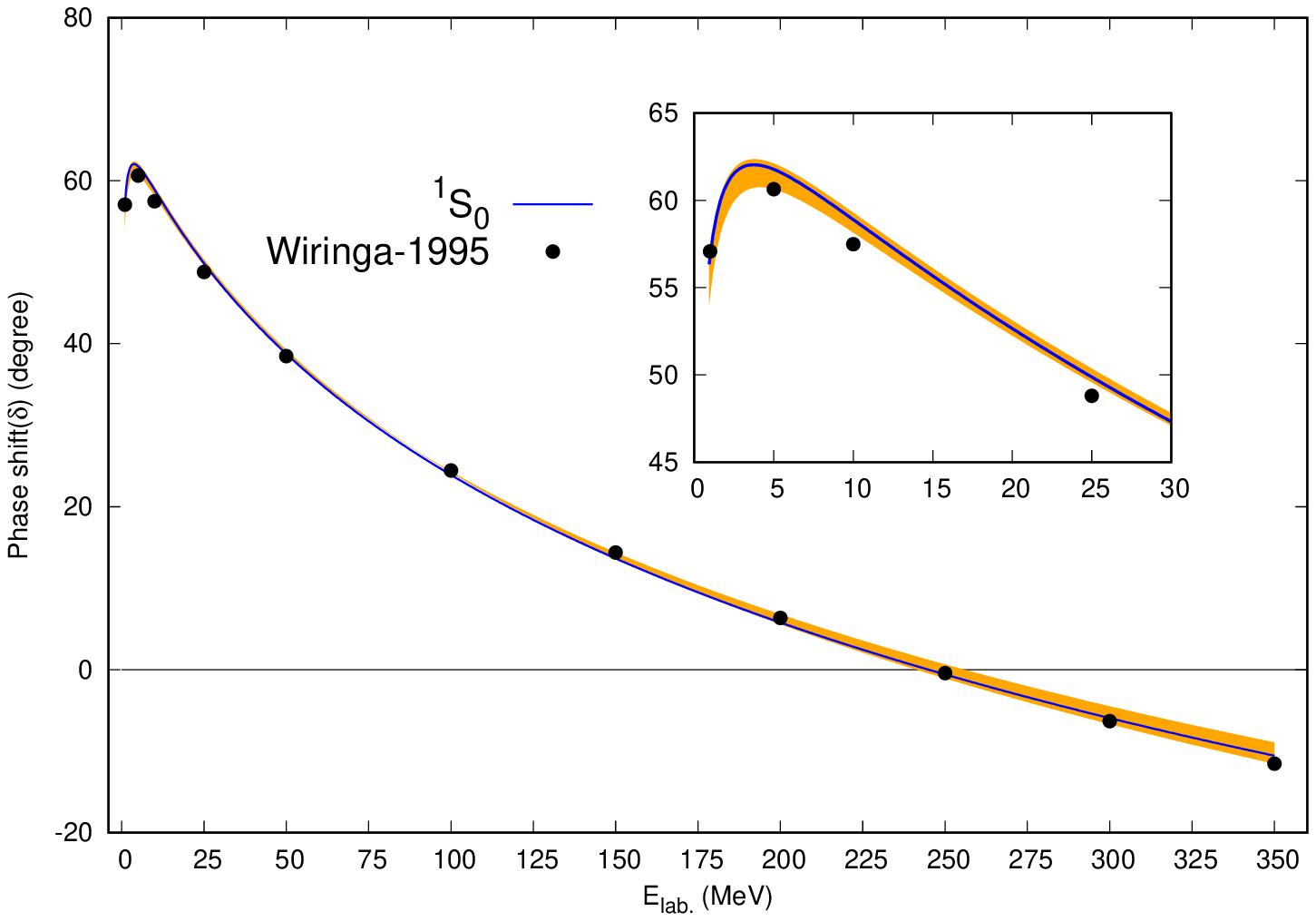} }
\caption{$^1S_0-nn$ SPS obtained through GOA (solid blue line) and CDA (yellow ribbon) analysis.}
\label{SPS_POT}
\end{figure*}

\clearpage
\section{Conclusion}
The outcomes of the paper is: \textit{nn} data from Wiringa \textit{et al.} have been fitted to obtain the class of interaction potentials which further produces scattering parameters ($r_e$ and \textit{a}) and pseudo bound state energy for the \textit{nn} system. For this purpose we used VMC as an optimization procedure which was ran in tandem with phase function method. CDA (computationally time taking procedure $\approx$ 2,20,000 iterations) was performed to overcome the overfitting nature of GOA (less computational time $\approx$ 500 iterations) such that the class of potentials generated via CDA could give rise to the potentials which are capable in obtaining the low energy scattering parameters and energy. In addition we also obtained quantum mechanical functions like: amplitude and wavefunctions for nn interaction. We will be communicating GOA plus CDA for higher states of \textit{nn} scattering in near future.  

\section{Appendix}\label{Appendix}
\subsection{Singlet State Analysis}
\begin{itemize}
\item Since, we have three parameters to be determined for $^1S_0-nn$ state, a total of 165 combinations need to considered. All these are shown in Table \ref{165combi}, where the data have been once again presented in ascending order of overall MAE.
\item Out of these 165, only 112 of them have $MAE < 2$ (Table \ref{165combi}). The average value for $r_m$ from these 112 combinations is determined to be 0.915 fm. 
\item Keeping $r_m=0.915$ fixed, one needs to vary only two parameter $V_0$ and $a_m$. So, only $^{11}C_2$, that is 55 combinations need to be worked out. These are given in Table \ref{55combi}.
\item A total of 21 combinations are having $MAE < 1$. These have been considered for determining energy scattering parameters ($a_s$ and $r_s$) and are shown in Table \ref{21combi}.
\item The values of depth $V_0$ and width $a_m$, given in bold (Table \ref{21combi}), are utilized for obtaining possible range of values for scattering parameters in our calculations.
\end{itemize}
{\small
\begin{longtable}{cccccccc}
\caption{\textbf{$^1S_0$ state:} Model parameters for 165 combinations, each with three lab energies and obtained by minimising MSE. The overall MAE is determined by obtaining SPS for remaining experimental data points. The data is sorted with respect to MAE in ascending order.}\\
\hline
\textbf{\small{Sr. No.}} & \multirow{2}{*}{\begin{tabular}[c]{@{}c@{}}$E_1$\\(MeV)\end{tabular}}  & \multirow{2}{*}{\begin{tabular}[c]{@{}c@{}}$E_2$\\(MeV)\end{tabular}}  & \multirow{2}{*}{\begin{tabular}[c]{@{}c@{}}$E_3$\\(MeV)\end{tabular}}  & \multirow{2}{*}{\begin{tabular}[c]{@{}c@{}}$V_0$\\(MeV)\end{tabular}}  & \multirow{2}{*}{\begin{tabular}[c]{@{}c@{}}$r_m$\\(fm)\end{tabular}} & \multirow{2}{*}{\begin{tabular}[c]{@{}c@{}}$a_m$\\(fm)\end{tabular}} & \multirow{2}{*}{\begin{tabular}[c]{@{}c@{}}\textbf{Overall}\\\textbf{MAE}\end{tabular}}  \\ 
\\
\hline
\endfirsthead
\multicolumn{8}{c}%
{\tablename\ \thetable\ -- \textit{Continued from previous page}} \\
\hline
\textbf{\small{Sr. No.}} & \multirow{2}{*}{\begin{tabular}[c]{@{}c@{}}$E_1$\\(MeV)\end{tabular}}  & \multirow{2}{*}{\begin{tabular}[c]{@{}c@{}}$E_2$\\(MeV)\end{tabular}}  & \multirow{2}{*}{\begin{tabular}[c]{@{}c@{}}$E_3$\\(MeV)\end{tabular}}  & \multirow{2}{*}{\begin{tabular}[c]{@{}c@{}}$V_0$\\(MeV)\end{tabular}}  & \multirow{2}{*}{\begin{tabular}[c]{@{}c@{}}$r_m$\\(fm)\end{tabular}} & \multirow{2}{*}{\begin{tabular}[c]{@{}c@{}}$a_m$\\(fm)\end{tabular}} & \multirow{2}{*}{\begin{tabular}[c]{@{}c@{}}\textbf{Overall}\\\textbf{MAE}\end{tabular}}  \\
\\ 
\hline
\endhead
\hline \multicolumn{8}{r}{\textit{Continued on next page}} \\
\endfoot
\hline
\endlastfoot
1	&5	&100	&300	&77.115479	&0.903579	&0.350886	&0.662538\\
2	&5	&100	&250	&74.794074	&0.905138	&0.356135	&0.664066\\
3	&5	&50	&250	&66.555057	&0.923181	&0.377144	&0.667124\\
4	&5	&50	&300	&68.714993	&0.923036	&0.371343	&0.667529\\
5	&1	&50	&250	&63.920909	&0.932580	&0.386325	&0.683027\\
6	&1	&50	&300	&66.096318	&0.932593	&0.380109	&0.685002\\
7	&1	&100	&250	&72.609854	&0.912592	&0.363168	&0.689700\\
8	&1	&100	&300	&74.974928	&0.911009	&0.357580	&0.689736\\
9	&5	&100	&350	&78.975352	&0.902434	&0.346849	&0.710647\\
10	&5	&150	&250	&80.031139	&0.895761	&0.344532	&0.724937\\
11	&5	&100	&200	&72.078202	&0.907161	&0.362599	&0.725451\\
12	&5	&150	&300	&82.482578	&0.893344	&0.339510	&0.734344\\
13	&1	&100	&350	&76.876625	&0.909845	&0.353275	&0.744264\\
14	&5	&150	&200	&77.282940	&0.898713	&0.350445	&0.744734\\
15	&5	&50	&350	&70.484158	&0.922960	&0.366796	&0.752458\\
16	&1	&100	&200	&69.825095	&0.914668	&0.370109	&0.756181\\
17	&1	&150	&250	&78.083707	&0.902422	&0.350603	&0.765884\\
18	&5	&150	&350	&84.381371	&0.891599	&0.335772	&0.770784\\
19	&5	&50	&200	&63.979704	&0.923428	&0.384456	&0.777142\\
20	&1	&150	&300	&80.592060	&0.899901	&0.345280	&0.778598\\
21	&1	&50	&350	&67.882315	&0.932638	&0.375236	&0.781104\\
22	&5	&200	&250	&83.695427	&0.889976	&0.337076	&0.792821\\
23	&5	&25	&250	&59.371657	&0.943254	&0.399038	&0.799728\\
24	&1	&50	&200	&61.317014	&0.932620	&0.394214	&0.805355\\
25	&5	&25	&300	&61.414424	&0.944445	&0.392522	&0.813973\\
26	&1	&150	&350	&82.547036	&0.898076	&0.341299	&0.820508\\
27	&5	&200	&300	&86.368368	&0.886791	&0.331951	&0.822296\\
28	&10	&50	&250	&69.079223	&0.914998	&0.369006	&0.826098\\
29	&10	&50	&300	&71.223951	&0.914683	&0.363555	&0.830384\\
30	&1	&25	&250	&56.293894	&0.955358	&0.410959	&0.840832\\
31	&1	&200	&250	&81.911322	&0.896193	&0.342571	&0.847136\\
32	&5	&200	&350	&88.277749	&0.884659	&0.328433	&0.858753\\
33	&1	&25	&300	&58.320941	&0.956947	&0.403929	&0.859694\\
34	&5	&100	&150	&68.510238	&0.910179	&0.371682	&0.867157\\
35	&10	&100	&250	&76.890502	&0.898528	&0.349814	&0.867544\\
36	&10	&100	&300	&79.172683	&0.896958	&0.344850	&0.872223\\
37	&10	&25	&250	&62.328957	&0.932878	&0.388595	&0.878832\\
38	&1	&200	&300	&84.642341	&0.892865	&0.337174	&0.882417\\
39	&10	&25	&300	&64.384332	&0.933691	&0.382510	&0.892251\\
40	&10	&50	&350	&72.976295	&0.914475	&0.359286	&0.909298\\
41	&1	&100	&150	&66.168294	&0.917781	&0.379889	&0.912405\\
42	&10	&100	&200	&74.240851	&0.900538	&0.355870	&0.914718\\
43	&5	&250	&300	&89.779662	&0.881551	&0.325726	&0.917400\\
44	&10	&50	&200	&66.533663	&0.915464	&0.375827	&0.918218\\
45	&10	&100	&350	&80.993022	&0.895804	&0.341044	&0.920199\\
46	&1	&200	&350	&86.617050	&0.890615	&0.333431	&0.924701\\
47	&10	&150	&250	&81.907492	&0.889782	&0.339023	&0.942778\\
48	&5	&25	&200	&56.917854	&0.941723	&0.407363	&0.944494\\
49	&5	&25	&350	&63.104300	&0.945385	&0.387386	&0.945559\\
50	&10	&150	&200	&79.250014	&0.892611	&0.344558	&0.952231\\
51	&5	&250	&350	&91.482999	&0.879424	&0.322753	&0.955824\\
52	&10	&150	&300	&84.307897	&0.887426	&0.334255	&0.958265\\
53	&1	&250	&300	&88.148662	&0.887337	&0.330608	&0.991374\\
54	&10	&150	&350	&86.153677	&0.885734	&0.330728	&0.996277\\
55	&1	&25	&200	&53.855529	&0.953255	&0.419992	&1.002329\\
56	&10	&25	&200	&59.867095	&0.931867	&0.396320	&1.004255\\
57	&10	&25	&350	&66.079670	&0.934349	&0.377716	&1.009266\\
58	&1	&25	&350	&60.000686	&0.958187	&0.398395	&1.010507\\
59	&10	&200	&250	&85.417081	&0.884354	&0.332064	&1.016014\\
60	&5	&300	&350	&93.552354	&0.876249	&0.319243	&1.027944\\
61	&10	&100	&150	&70.755154	&0.903526	&0.364365	&1.029546\\
62	&1	&250	&350	&89.938622	&0.885053	&0.327404	&1.036422\\
63	&5	&50	&150	&60.654246	&0.923860	&0.394616	&1.046363\\
64	&10	&200	&300	&88.041188	&0.881264	&0.327161	&1.050844\\
65	&10	&200	&350	&89.885755	&0.879225	&0.323846	&1.088957\\
66	&5	&10	&250	&52.061359	&0.969479	&0.426049	&1.092622\\
67	&1	&50	&150	&57.955534	&0.932734	&0.405222	&1.108618\\
68	&1	&300	&350	&92.123306	&0.881627	&0.323616	&1.120125\\
69	&5	&10	&300	&53.988044	&0.972202	&0.418513	&1.130341\\
70	&10	&50	&150	&63.245032	&0.916219	&0.385275	&1.147237\\
71	&10	&250	&300	&91.366002	&0.876248	&0.321223	&1.148571\\
72	&10	&250	&350	&92.984555	&0.874252	&0.318459	&1.187414\\
73	&1	&10	&250	&48.534912	&0.985705	&0.441887	&1.226465\\
74	&10	&300	&350	&94.940727	&0.871296	&0.315198	&1.257704\\
75	&5	&10	&200	&49.742061	&0.965777	&0.435794	&1.269821\\
76	&1	&10	&300	&50.407365	&0.989203	&0.433686	&1.275819\\
77	&25	&50	&250	&74.165885	&0.900566	&0.354256	&1.287763\\
78	&25	&50	&300	&76.288280	&0.899839	&0.349376	&1.308031\\
79	&5	&10	&350	&55.591458	&0.974292	&0.412578	&1.323138\\
80	&5	&25	&150	&53.769108	&0.939484	&0.418971	&1.325565\\
81	&10	&25	&150	&56.704177	&0.930439	&0.407046	&1.333711\\
82	&25	&50	&200	&71.675838	&0.901540	&0.360268	&1.336092\\
83	&25	&50	&350	&78.008622	&0.899316	&0.345574	&1.382325\\
84	&1	&10	&200	&46.281762	&0.980843	&0.452557	&1.422110\\
85	&1	&150	&200	&101.253374	&0.880595	&0.309167	&1.424588\\
86	&1	&25	&150	&50.732258	&0.950087	&0.432647	&1.434219\\
87	&25	&50	&150	&68.456491	&0.903012	&0.368552	&1.476343\\
88	&25	&100	&250	&81.169693	&0.886459	&0.338039	&1.499279\\
89	&1	&10	&350	&51.971163	&0.991852	&0.427233	&1.500739\\
90	&25	&100	&200	&78.640776	&0.888565	&0.343416	&1.503322\\
91	&25	&100	&300	&83.385047	&0.884770	&0.333542	&1.529394\\
92	&1	&5	&250	&44.703470	&1.004339	&0.460146	&1.531274\\
93	&25	&100	&150	&75.299148	&0.891671	&0.350961	&1.544265\\
94	&5	&50	&100	&56.161277	&0.924590	&0.409879	&1.566059\\
95	&10	&50	&100	&58.818554	&0.917491	&0.399325	&1.586276\\
96	&25	&100	&350	&85.132502	&0.883530	&0.330126	&1.587127\\
97	&1	&5	&300	&46.491365	&1.008976	&0.451198	&1.600957\\
98	&25	&150	&200	&83.291825	&0.881340	&0.333479	&1.622311\\
99	&25	&150	&250	&85.776189	&0.878634	&0.328588	&1.651279\\
100	&25	&150	&300	&88.087350	&0.876299	&0.324237	&1.695171\\
101	&1	&50	&100	&53.407413	&0.932875	&0.421910	&1.701221\\
102	&1	&5	&200	&42.555702	&0.997748	&0.471868	&1.742130\\
103	&25	&50	&100	&64.168570	&0.905385	&0.380620	&1.746915\\
104	&25	&150	&350	&89.830087	&0.874644	&0.321074	&1.747854\\
105	&25	&200	&250	&88.985575	&0.873745	&0.322482	&1.765219\\
106	&5	&10	&150	&46.786039	&0.960056	&0.449510	&1.772808\\
107	&25	&200	&300	&91.528935	&0.870731	&0.317930	&1.830865\\
108	&1	&5	&350	&47.990655	&1.012439	&0.444169	&1.870402\\
109	&25	&200	&350	&93.247202	&0.868811	&0.314968	&1.884770\\
110	&25	&250	&300	&94.707257	&0.866003	&0.312440	&1.963546\\
111	&1	&10	&150	&43.425162	&0.973119	&0.467661	&1.991153\\
112	&10	&25	&100	&52.429141	&0.928047	&0.423295	&\textbf{1.997752}\\
\hline
113	&25	&250	&350	&96.143870	&0.864233	&0.310072	&2.013517\\
114	&25	&300	&350	&97.857572	&0.861669	&0.307287	&2.094282\\
115	&5	&25	&100	&49.514747	&0.935561	&0.436721	&2.101010\\
116	&50	&100	&250	&85.638400	&0.875559	&0.327097	&2.166302\\
117	&50	&100	&300	&87.817526	&0.873615	&0.322932	&2.232250\\
118	&50	&100	&350	&89.496261	&0.872214	&0.319838	&2.306645\\
119	&1	&25	&100	&46.523119	&0.944299	&0.452180	&2.322517\\
120	&50	&150	&200	&87.586643	&0.870942	&0.322978	&2.354390\\
121	&1	&5	&150	&39.854431	&0.986974	&0.488576	&2.381987\\
122	&50	&150	&250	&89.892435	&0.868248	&0.318643	&2.433698\\
123	&50	&150	&300	&92.144370	&0.865792	&0.314587	&2.520239\\
124	&50	&150	&350	&93.782322	&0.864105	&0.311741	&2.595768\\
125	&50	&200	&250	&92.796838	&0.863730	&0.313268	&2.609630\\
126	&50	&200	&300	&95.310954	&0.860607	&0.308912	&2.726528\\
127	&50	&200	&350	&96.894458	&0.858748	&0.306268	&2.804948\\
128	&5	&10	&100	&42.834193	&0.949276	&0.470912	&2.835505\\
129	&50	&250	&300	&98.394826	&0.855944	&0.303696	&2.923710\\
130	&50	&250	&350	&99.610614	&0.854397	&0.301742	&2.988923\\
131	&50	&300	&350	&101.038510	&0.852221	&0.299448	&3.086309\\
132	&1	&10	&100	&39.645644	&0.957950	&0.491504	&3.204863\\
133	&100	&150	&200	&93.222892	&0.859267	&0.310799	&3.215942\\
134	&100	&150	&250	&95.213565	&0.856649	&0.307229	&3.350851\\
135	&10	&25	&50	&45.667704	&0.921380	&0.454834	&3.443735\\
136	&100	&150	&300	&97.445560	&0.853883	&0.303385	&3.501372\\
137	&100	&200	&250	&97.627765	&0.852638	&0.302836	&3.597631\\
138	&100	&150	&350	&98.929260	&0.852136	&0.300916	&3.605990\\
139	&1	&5	&100	&36.346508	&0.964850	&0.515314	&3.763530\\
140	&100	&200	&300	&100.240006	&0.849066	&0.298428	&3.805282\\
141	&5	&25	&50	&42.830270	&0.924177	&0.471778	&3.810106\\
142	&100	&200	&350	&101.632354	&0.847259	&0.296164	&3.915039\\
143	&150	&200	&250	&100.455374	&0.846841	&0.297276	&4.110838\\
144	&100	&250	&300	&103.385872	&0.844043	&0.293166	&4.124725\\
145	&100	&250	&350	&104.197584	&0.842930	&0.291885	&4.193478\\
146	&100	&300	&350	&105.126331	&0.841428	&0.290387	&4.291667\\
147	&1	&25	&50	&39.988895	&0.926528	&0.491485	&4.297531\\
148	&150	&200	&300	&103.488161	&0.842317	&0.292178	&4.430524\\
149	&150	&200	&350	&104.716717	&0.840585	&0.290197	&4.554876\\
150	&150	&250	&300	&107.091983	&0.836250	&0.286146	&4.904307\\
151	&150	&300	&350	&107.712611	&0.835260	&0.285151	&4.981076\\
152	&250	&300	&350	&108.438873	&0.833609	&0.283742	&5.165724\\
153	&5	&10	&50	&36.900274	&0.914826	&0.515154	&5.246407\\
154	&200	&300	&350	&109.443045	&0.831385	&0.281837	&5.412344\\
155	&200	&250	&350	&110.400050	&0.829656	&0.280247	&5.569773\\
156	&200	&250	&300	&111.292402	&0.828317	&0.278881	&5.680566\\
157	&1	&10	&50	&34.242821	&0.905331	&0.542148	&5.964818\\
158	&1	&5	&50	&31.855840	&0.880018	&0.574156	&6.904728\\
159	&5	&10	&25	&33.160455	&0.827556	&0.571797	&7.980513\\
160	&1	&10	&25	&32.404745	&0.749428	&0.610644	&9.053548\\
161	&1	&5	&25	&35.401342	&0.554285	&0.660232	&10.338711\\
162	&1	&5	&10	&60.167715	&0.010001	&0.707299	&11.760083\\
163	&50	&100	&200	&21.505459	&2.000000	&1.357916	&19.074560\\
164	&50	&100	&150	&18.792839	&2.000000	&1.460051	&19.352376\\
165	&150	&250	&350	&39.237156	&2.000000	&1.232217	&25.195640\\
\label{165combi}
\end{longtable}
}

{\small
\begin{longtable}{cccccccc}
\caption{\textbf{$^1S_0$ state:} Model parameters for 55 combinations for $^1S_0-nn$ state. After fixing $r_m=0.9155$ (average) fm from 112 combinations of previous table \ref{165combi}, two parameters produces $^{11}C_2$ \textit{i.e.} 55 combinations. The data is sorted with respect to MAE in ascending order.}\\
\hline
\textbf{\small{Sr. No.}} & \multirow{2}{*}{\begin{tabular}[c]{@{}c@{}}$E_1$\\(MeV)\end{tabular}} & \multirow{2}{*}{\begin{tabular}[c]{@{}c@{}}$E_2$\\(MeV)\end{tabular}} & \multirow{2}{*}{\begin{tabular}[c]{@{}c@{}}$V_0$\\(MeV)\end{tabular}}  & \multirow{2}{*}{\begin{tabular}[c]{@{}c@{}}$a_m$\\(fm)\end{tabular}} & \multirow{2}{*}{\begin{tabular}[c]{@{}c@{}}\textbf{Overall}\\\textbf{MAE}\end{tabular}}  \\ 
\\
\hline
\endfirsthead
\multicolumn{6}{c}%
{\tablename\ \thetable\ -- \textit{Continued from previous page}} \\
\hline
\textbf{\small{Sr. No.}} & \multirow{2}{*}{\begin{tabular}[c]{@{}c@{}}$E_1$\\(MeV)\end{tabular}} & \multirow{2}{*}{\begin{tabular}[c]{@{}c@{}}$E_2$\\(MeV)\end{tabular}} & \multirow{2}{*}{\begin{tabular}[c]{@{}c@{}}$V_0$\\(MeV)\end{tabular}}  & \multirow{2}{*}{\begin{tabular}[c]{@{}c@{}}$a_m$\\(fm)\end{tabular}} & \multirow{2}{*}{\begin{tabular}[c]{@{}c@{}}\textbf{Overall}\\\textbf{MAE}\end{tabular}}  \\ 
\\ 
\hline
\endhead
\hline \multicolumn{8}{r}{\textit{Continued on next page}} \\
\endfoot
\hline
\endlastfoot
1	&5	&300	&71.730802	&0.363578	&0.666437	&\\
2	&5	&250	&69.812442	&0.368384	&0.666720	&\\
3	&1	&250	&71.193511	&0.366651	&0.689214	&\\
4	&1	&300	&72.930120	&0.362403	&0.689361	&\\
5	&5	&50	&72.914193	&0.360607	&0.736652	&\\
6	&5	&350	&73.343871	&0.359686	&0.738354	&\\
7	&5	&200	&67.660861	&0.374021	&0.753482	&\\
8	&1	&350	&74.427456	&0.358861	&0.754361	&\\
9	&1	&200	&69.374385	&0.371273	&0.758771	&\\
10	&50	&300	&70.968527	&0.364323	&0.814791	&\\
11	&50	&250	&68.917538	&0.369509	&0.816551	&\\
12	&100	&200	&69.587353	&0.370934	&0.817602	&\\
13	&10	&250	&68.868276	&0.369571	&0.827726	&\\
14	&10	&300	&70.898325	&0.364391	&0.833286	&\\
15	&50	&350	&72.663694	&0.360200	&0.890809	&\\
16	&100	&150	&66.847782	&0.377443	&0.900150	&\\
17	&100	&250	&71.804840	&0.365886	&0.911899	&\\
18	&10	&350	&72.583883	&0.360260	&0.914793	&\\
19	&50	&200	&66.521411	&0.375867	&0.917600	&\\
20	&10	&200	&66.517146	&0.375874	&0.918413	&\\
21	&5	&150	&65.191362	&0.380843	&\textbf{0.939942}	&\\
\hline
22	&100	&300	&73.759048	&0.361589	&1.044861	&\\
23	&25	&250	&68.110429	&0.370526	&1.104946	&\\
24	&25	&300	&70.216211	&0.365056	&1.122282	&\\
25	&10	&150	&63.620993	&0.384135	&1.141107	&\\
26	&5	&100	&63.240549	&0.386527	&1.142037	&\\
27	&50	&150	&63.503218	&0.384385	&1.165667	&\\
28	&25	&200	&65.625739	&0.377329	&1.190309	&\\
29	&100	&350	&75.376753	&0.358133	&1.206715	&\\
30	&25	&350	&71.951289	&0.360735	&1.217653	&\\
31	&25	&150	&62.456573	&0.386623	&1.421515	&\\
32	&50	&100	&59.621094	&0.396317	&1.614747	&\\
33	&150	&200	&72.531742	&0.366298	&1.691748	&\\
34	&25	&100	&58.241976	&0.400252	&1.876912	&\\
35	&150	&250	&74.641661	&0.362351	&2.037006	&\\
36	&150	&300	&76.578661	&0.358811	&2.376252	&\\
37	&150	&350	&78.165101	&0.355970	&2.684166	&\\
38	&10	&50	&101.859853	&0.305723	&2.942245	&\\
39	&200	&250	&77.056751	&0.359363	&3.026641	&\\
40	&200	&300	&79.074761	&0.356343	&3.612583	&\\
41	&200	&350	&80.655030	&0.354009	&4.083227	&\\
42	&10	&25	&42.587021	&0.472826	&4.263083	&\\
43	&5	&25	&40.200809	&0.489460	&4.659452	&\\
44	&250	&300	&81.441998	&0.354000	&4.800802	&\\
45	&1	&25	&37.934492	&0.507781	&5.094683	&\\
46	&5	&10	&36.968052	&0.514500	&5.213176	&\\
47	&1	&50	&36.061249	&0.523395	&5.384020	&\\
48	&250	&350	&82.929463	&0.352199	&5.389926	&\\
49	&1	&10	&34.833315	&0.534637	&5.588524	&\\
50	&1	&5	&32.866784	&0.554584	&5.952004	&\\
51	&300	&350	&84.672221	&0.350802	&6.390542	&\\
52	&25	&50	&30.513851	&0.596846	&7.700140	&\\
53	&1	&100	&21.346168	&0.752283	&7.851272	&\\
54	&10	&100	&18.349456	&0.917733	&9.809530	&\\
55	&1	&150	&10.445956	&1.341045	&9.942812	&\\
\label{55combi}
\end{longtable}
}

{\small
\begin{longtable}{cccccccc}
\caption{\textbf{$^1S_0$ state:} Model parameters for 21 combinations, each with three lab energies and obtained by minimising MSE. The overall MAE is determined by obtaining SPS for remaining experimental data points. The data is sorted with respect to $E_1$ in ascending order.}\\
\hline
\textbf{\small{Sr. No.}} & \multirow{2}{*}{\begin{tabular}[c]{@{}c@{}}$E_1$\\(MeV)\end{tabular}}  & \multirow{2}{*}{\begin{tabular}[c]{@{}c@{}}$E_2$\\(MeV)\end{tabular}}  & \multirow{2}{*}{\begin{tabular}[c]{@{}c@{}}$V_0$\\(MeV)\end{tabular}}  & \multirow{2}{*}{\begin{tabular}[c]{@{}c@{}}$a_m$\\(MeV)\end{tabular}}  & \multirow{2}{*}{\begin{tabular}[c]{@{}c@{}}\textbf{Overall}\\\textbf{MAE}\end{tabular}}& \multirow{2}{*}{\begin{tabular}[c]{@{}c@{}}$r_e$\\(fm)\end{tabular}} & \multirow{2}{*}{\begin{tabular}[c]{@{}c@{}}$a$\\(fm)\end{tabular}}   \\
\\
\hline
\endfirsthead
\multicolumn{8}{c}%
{\tablename\ \thetable\ -- \textit{Continued from previous page}} \\
\hline
\textbf{\small{Sr. No.}} & \multirow{2}{*}{\begin{tabular}[c]{@{}c@{}}$E_1$\\(MeV)\end{tabular}}  & \multirow{2}{*}{\begin{tabular}[c]{@{}c@{}}$E_2$\\(MeV)\end{tabular}}  & \multirow{2}{*}{\begin{tabular}[c]{@{}c@{}}$V_0$\\(MeV)\end{tabular}}  & \multirow{2}{*}{\begin{tabular}[c]{@{}c@{}}$a_m$\\(MeV)\end{tabular}}  & \multirow{2}{*}{\begin{tabular}[c]{@{}c@{}}\textbf{Overall}\\\textbf{MAE}\end{tabular}}& \multirow{2}{*}{\begin{tabular}[c]{@{}c@{}}$r_e$\\(fm)\end{tabular}} & \multirow{2}{*}{\begin{tabular}[c]{@{}c@{}}$a$\\(fm)\end{tabular}}   \\
\\ 
\hline
\endhead
\hline \multicolumn{8}{r}{\textit{Continued on next page}} \\
\endfoot
\hline
\endlastfoot
1	&1	&250	&71.193511	&0.366651	&0.689214	&-18.380	&2.539\\
2	&1	&300	&72.930120	&0.362403	&0.689361	&-18.373	&2.534\\
3	&1	&350	&\textbf{74.427456}	&\textbf{0.358861}	&0.754361	&-18.368	&2.530\\
4	&1	&200	&69.374385	&0.371273	&0.758771	&-18.387	&2.544\\
5	&5	&300	&71.730802	&0.363578	&0.666437	&-16.173	&2.556\\
6	&5	&250	&69.812442	&0.368384	&0.666720	&-16.212	&2.561\\
7	&5	&50	&72.914193	&0.360607	&0.736652	&-16.076	&2.554\\
8	&5	&350	&73.343871	&0.359686	&0.738354	&-16.142	&2.552\\
9	&5	&200	&67.660861	&0.374021	&0.753482	&-16.258	&2.566\\
10	&5	&150	&\textbf{65.191362}	&\textbf{0.380843}	&0.939942	&-16.314	&2.573\\
11	&10	&250	&68.868276	&0.369571	&0.827726	&-14.962	&2.576\\
12	&10	&300	&70.898325	&0.364391	&0.833286	&-14.896	&2.571\\
13	&10	&350	&72.583883	&0.360260	&0.914793	&-14.844	&2.568\\
14	&10	&200	&66.517146	&0.375874	&0.918413	&-15.043	&2.582\\
15	&50	&300	&70.968527	&0.364323	&0.814791	&-14.997	&2.570\\
16	&50	&250	&68.917538	&0.369509	&0.816551	&-15.024	&2.575\\
17	&50	&350	&72.663694	&0.360200	&0.890809	&-14.972	&2.566\\
18	&50	&200	&66.521411	&0.375867	&0.917600	&-15.047	&2.582\\
19	&100	&200	&69.587353	&0.370934	&0.817602	&-18.681	&2.541\\
20	&100	&150	&66.847782	&0.377443	&0.900150	&-17.673	&2.556\\
21	&100	&250	&71.804840	&0.365886	&0.911899	&-19.499	&2.529\\
\hline
	&	&Avg.	&70.226561	&0.367646	&0.807472	&-16.492	&2.558\\
	&	&St. Dev.	&2.605663	&0.006510	&0.091322	&1.557	&0.017\\
\label{21combi}
\end{longtable}
}

\end{document}